\documentclass[reprint,amsmath,amssymb,aps,prl,superscriptaddress, showpacs, eprint]{revtex4-2}
\usepackage[utf8]{inputenc}

\usepackage{amsmath,amssymb,amsthm,ascmac,fancybox}
\usepackage{color}
\usepackage{graphicx}
\usepackage{url}
\usepackage{siunitx}
\usepackage{bm}
\usepackage{physics}
\usepackage{mathtools}
\usepackage{hyperref}
\usepackage[T1]{fontenc}
\usepackage{ulem}
\usepackage{docmute} %

\graphicspath{figures/}

\newcommand{\ene}{\varepsilon}
\renewcommand{\vec}{\vb*}

\AtBeginDocument{%
  \renewcommand{\selectlanguage}[1]{}%
}

\usepackage{multirow}
\usepackage{booktabs}

\begin{document}
\title{Singular high-harmonic transport above a quantum threshold}
\author{Yugo Onishi}
\altaffiliation{Current address: Department of Physics, Stanford University, Stanford, CA 94305}
\affiliation{Department of Physics, Massachusetts Institute of Technology, Cambridge, MA 02139, USA}

\author{Su-Yang Xu}
\affiliation{Department of Chemistry and Chemical Biology, Harvard University, Cambridge, MA, USA}
\author{Liang Fu}
\affiliation{Department of Physics, Massachusetts Institute of Technology, Cambridge, MA 02139, USA}
\date{\today}

\begin{abstract}
    We show that Landau-Zener tunneling in a small-gap insulator produces a singular DC current--voltage relation, and correspondingly, strong high-harmonic generation in AC transport. Remarkably, high-harmonic responses decrease subexponentially with harmonic order, which is parametrically slower than in ordinary conductors and PN diodes.
    We further derive a scaling law for the current amplitude at frequency $n\omega$ produced by the applied electric field $E$ at fundamental frequency $\omega$. 
    Our work offers a promising route to THz generation based on frequency multiplication in solids. 
\end{abstract}

\maketitle

\textit{Introduction. ---} DC electrical conduction requires mobile carriers. In metals, an applied electric field accelerates carriers near the Fermi surface. In a band insulator, by contrast, the absence of a Fermi surface and the finite energy gap forbids DC conduction within linear response.

A sufficiently strong electric field changes this situation qualitatively: electrons can tunnel from valence to conduction bands, creating electron and holes, giving rise to a finite current~\cite{zener1934,kaneZenerTunnelingSemiconductors1960,landauQuantumMechanicsNonrelativistic1977}. Because the field generates carriers in addition to accelerating carriers, strong-field transport in insulators is intrinsically nonlinear and nonperturbative~\cite{kitamuraNonreciprocalLandauZener2020,kitamuraCurrentResponseNonequilibrium2020}.

This mechanism differs from perturbative nonlinear responses in metals and semimetals~\cite{rikken2005,Pop2014,Sodemann2015,gao2014,ma2019a,gao2023,wang2023,Tokura2018,Ideue2017,lai2021,min2023a,chou2025a,orensteinTopologySymmetryQuantum2021,isobe2020,zhang2018a,Watanabe2020a}, for which the current is expanded as $J=\sigma E+\chi E^2+\cdots$. In a clean band insulator under a DC field, the leading current instead arises from interband tunneling and cannot be captured by any finite-order expansion around equilibrium. The natural question is then: what is the distinctive experimental feature of the nonperturbative strong-field transport in insulators?

In this work, we identify the qualitatively distinct characteristics of highly nonlinear transport due to interband tunneling, which we call \textit{singular high-harmonic transport}. 
We first show that the tunneling produces a sharp onset of conduction, resulting in a singularity in the current--voltage relation. When driven by an AC field with amplitude $E_0$, the singularity produces strong high harmonics: at large odd harmonic order $n$, the amplitude decays as $J_n\sim\exp(-\order{\sqrt{n}})$. This is much slower than in other systems without a singularity, such as an ideal PN diode, and hence the singularity governs the high-harmonic behavior. Furthermore, as a key feature of the singular high-harmonic transport, we identify a scaling law that relates $J_n(E_0)$ for different $n$: plots of $(-1)^{(n+1)/2}nJ_n$ vs $E_0/n$ for different $n$ collapse onto one curve at large $n$. 

The singular high-harmonic transport is expected to appear generically in semiconductors and insulators, over a broad frequency range from quasi-DC to sub-THz, offering a new way to generate THz waves. It can occur even at relatively small electric fields $\sim \SI{1}{kV/m}$ in systems with a small or vanishing gap, such as (gated) graphene, Weyl/Dirac semimetals, systems near topological phase transitions, and the surface states of topological insulators.

\textit{Current-voltage relation. ---}
We consider an insulator with a small gap near a particular point in momentum space.
For simplicity, we first consider a one-dimensional system and discuss higher dimensions later. Its low-energy physics is described by the $k\vdot p$ Hamiltonian:
\begin{align}
    H(k) = \hbar v k \sigma_z + \Delta \sigma_x, \label{eq:1d_dirac}
\end{align}
where $k$ is the wavevector, $\Delta$ is the Dirac mass (half the band gap), and $v$ is the velocity. The dispersion is $\ene_{\pm}(k)=\pm\sqrt{\Delta^2 + \hbar^2 v^2 k^2}$. We assume zero temperature and a chemical potential in the gap, so that no carriers are present in equilibrium.

Under a DC electric field $E$, the crystal momentum evolves according to
    $\hbar\dot{k} = e E$.
Thus an electron with initial wavevector $k_0$ at time $t_0$ evolves under $H(k(t))$, where $k(t) = k_0 + (eE/\hbar)(t-t_0)$. 

A strong applied field can generate carriers by inducing tunneling from the valence band to the conduction band. Because $H(k(t))$ is the Landau-Zener Hamiltonian~\cite{zener1934,landauQuantumMechanicsNonrelativistic1977}, an electron prepared in the valence band at $k\to-\infty$
tunnels to the conduction band at $k=+\infty$ with probability
\begin{align}
    &P_{\rm LZ}(E) = \exp(-\frac{\pi\Delta^2}{\hbar |ve E|}) = e^{-E_c/|E|}, \label{eq:LZ_prob} \\
    &E_c = \frac{\pi\Delta^2}{\hbar ve},
\end{align}
where $E_c$ is the characteristic field at which tunneling becomes appreciable.

We observe that the probability~\eqref{eq:LZ_prob} has a simple interpretation. The combination $veE$ is the rate at which the field does work on a carrier moving with velocity $v$. Eq.~\eqref{eq:LZ_prob} therefore identifies the characteristic quantum power scale
\begin{align}
    w_{Q} &= \Delta^2/\hbar,
\end{align}
such that $veE_c=\pi w_Q$. $w_Q$ represents the power threshold at which the electrons overcome the energy gap by tunneling and start to generate carriers. 
We refer to $w_Q$ and the associated field $E_c$ as the \textit{quantum power threshold}.

\begin{figure}
    \centering
	\includegraphics[width=1.0\linewidth]{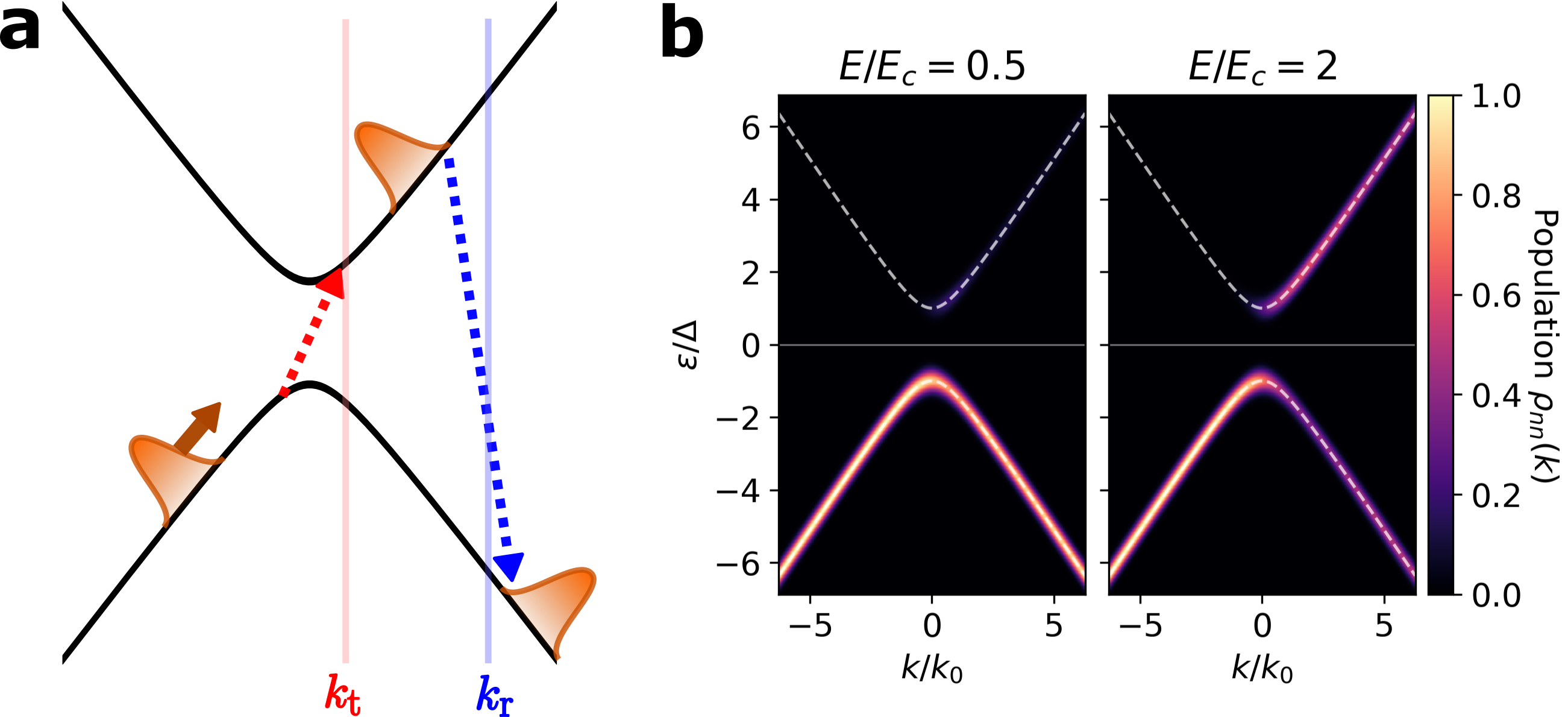}
    \caption{Landau-Zener tunneling and relaxation. (a) Electrons tunnel from the valence band to the conduction band within $|k|\lesssim k_{\rm t}$ and relax over the scale $|k|\sim k_{\rm r}$. (b) Electron population at finite field calculated with Eq.~\eqref{eq:g_Boltzmann}; dashed lines show the band dispersion. $k_0=\Delta/(\hbar v)$. The relaxation time is $\tau=5.0\hbar/\Delta$.}
    \label{fig:LZ_tunneling}
\end{figure}

Note that $P_{\rm LZ}(E)$ exhibits a sharp turn-on behavior near the threshold $E_c$. This behavior is mathematically characterized by its essential singularity at $E=0$, and cannot be captured by any finite-order expansion in $E$. This behavior determines the high-order harmonic transport discussed below.

We first calculate the DC current--voltage relation. To discuss the DC transport, we need to include the relaxation effect to obtain a steady state. 
To this end, we use the quantum master equation (also called semiconductor Bloch equation) for the one-body reduced density matrix $\rho(k,t)$ with relaxation time approximation~\cite{meierCoherentElectricFieldEffects1994,satoLightinducedAnomalousHall2019}:
\begin{align}
    \pdv{\rho}{t} + \dot{k}\pdv{\rho}{k} = -\frac{i}{\hbar}\comm{H(k)}{\rho} - \frac{\rho-\rho_0}{\tau}, \label{eq:g_Boltzmann}
\end{align}
where $\rho_0$ is the one-body reduced density matrix at equilibrium, $\dot{k}=eE/\hbar$, and $\tau$ is the relaxation time. For simplicity, we assume the same relaxation time for carrier relaxation (diagonal) and dephasing (off-diagonal). We assume $\tau$ is long so that the band gap is well-defined, i.e.,
\begin{align}
    \tau \gg \hbar/\Delta. \label{eq:large_tau}
\end{align}
Eq.~\eqref{eq:g_Boltzmann} can be viewed as a generalization of the semiclassical Boltzmann equation on a distribution function to an equation on a density matrix. 
We focus on $eE>0$, as the $eE<0$ case follows from inversion symmetry.

For a steady state with $\pdv*{\rho}{t}=0$ under a finite $E$, we can rewrite Eq.~\eqref{eq:g_Boltzmann} as 
\begin{align}
    \pdv{\rho}{k} = -\frac{i}{eE}\comm{H(k)}{\rho} - \frac{\rho-\rho_0}{eE\tau/\hbar}. \label{eq:g_Boltzmann2}
\end{align}
This equation consists of two parts. The first is the commutator on the right-hand side, which describes the carrier generation due to the interband tunneling. It is in the same form as the equation of motion for the reduced density matrix with a rescaled Hamiltonian $H(k)/(eE)$ by regarding the wavevector $k$ as the time $t$. The second is the relaxation term, which describes the decay of the generated carrier population through electron-hole recombination. 

In the following, we solve this equation approximately and calculate the steady-state current. Two momentum scales in the problem play important roles: the scale for interband tunneling $k_{\rm t}$, and the scale for relaxation $k_{\rm r}$, both defined below. We will show that carrier generation by interband tunneling and decay by the relaxation process jointly determine the occupation of conduction and valence band states at large $k$. In particular, 
$k_{\rm t}\ll k_{\rm r}$ holds in the regime of interest, and this separation of scales allows us to construct a good approximate solution.

Let us first define the scale for the tunneling $k_{\rm t}$. This is the momentum scale over which most of the tunneling takes place in the clean limit. 
In the clean limit $\tau = \infty$, Eq.~\eqref{eq:g_Boltzmann2} reduces to the Landau-Zener problem, and for the boundary condition $\rho(k\to-\infty)=\rho_0$ (assuming electric field $eE>0$), the conduction- and valence-band populations approximately satisfy~\cite{mullenTimeZenerTunneling1989c,vitanovTransitionTimesLandauZener1999,zenesiniTimeResolvedMeasurementLandauZener2009,shevchenko2010}
\begin{align}
    \delta\rho_{cc}(k) = -\delta\rho_{vv}(k) \approx
    \begin{cases}
        0 \quad (k \lesssim -k_{\rm t}) \\
        P_{\rm LZ}(E) \quad (k\gtrsim k_{\rm t})
    \end{cases} (\tau=\infty)\label{eq:clean_finite_k}
\end{align}
where
\begin{align}
    k_{\rm t} &=  \frac{\Delta}{2\hbar v}\mathrm{max}(1, \sqrt{2|E|/E_c}) \label{eq:kt}
\end{align}
For $|E|\le E_c/2$, $k_{\rm t}$ describes the momentum-space region of strong band hybridization and is independent of $E$, while for $|E|\ge E_c/2$, $k_{\rm t}=\sqrt{|eE|/(2\pi\hbar v)}$ is independent of $\Delta$. Note that in the absence of relaxation ($\tau=\infty$), the carrier distribution tail at large momentum generated by interband tunneling is $k$-independent.

The momentum scale for the relaxation is defined as
\begin{align}
    k_{\rm r} = |eE|\tau/\hbar. \label{eq:kr}
\end{align}
$k_{\rm r}$ is the change of momentum acquired from the electric field before an electron relaxes. 
In the absence of interband tunneling,  
Eq.~\eqref{eq:g_Boltzmann2} admits the following solution:
\begin{align}
    \delta\rho \propto \exp(-k/k_{\rm r}) \quad (\mathrm{finite\ } \tau) \label{eq:exp_damping}
\end{align}
Relaxation is negligible on momentum scales much smaller than $k_{\rm r}$.

Importantly, the two scales are well separated as $k_{\rm t}\ll k_{\rm r}$ when the following condition holds:
\begin{align}
    |E|/E_c \gg \hbar/(\tau \Delta). \label{eq:E_cond}
\end{align}
Given Eq.~\eqref{eq:large_tau}, this condition %
holds throughout the onset regime $|E|\sim E_c$.

Under the condition~\eqref{eq:E_cond} and hence the scale separation, tunneling occurs as in the clean limit (Eq.~\eqref{eq:clean_finite_k}) for $|k|\lesssim k_{\rm t}\ll k_{\rm r}$ and generates carriers, whereas the generated carriers relaxes exponentially over $k_{\rm r}$ as Eq.~\eqref{eq:exp_damping}. Thus, for $|k|\gtrsim k_{\rm t}$,
\begin{align}
    \delta\rho_{cc}(k) = -\delta\rho_{vv}(k) \approx 
    \begin{cases}
        0 \quad (k \lesssim -k_{\rm t}) \\
        P_{\rm LZ}(E)e^{-k/k_{\rm r}} \quad (k\gtrsim k_{\rm t})
    \end{cases} \label{eq:dist_approx}
\end{align}
Up to a relative error of $\order{k_{\rm t}/k_{\rm r}}$, we can further approximate Eq.~\eqref{eq:dist_approx} as
\begin{align}
    \delta\rho_{cc}(k) = -\delta\rho_{vv}(k) \approx \Theta(k) P_{\rm LZ}(E)e^{-k/k_{\rm r}} \label{eq:dist}
\end{align}

The current can be calculated with the current operator $J=ev\sigma_z$. Note that at large $k\gg k_{\rm t}$, the conduction- and valence-band states are approximately eigenstates of $\sigma_z$, and thus the current contribution from large $k$ is determined approximately by $\rho_{cc}-\rho_{vv}$. In addition, since the dominant current comes from $k\gg k_{\rm t}$ when $k_{\rm r}\gg k_{\rm t}$, we can approximate the current density as  $J\approx ev\int \dd{k}\,[\delta\rho_{cc}(k)-\delta\rho_{vv}(k)]/(2\pi)$, which yields
\begin{align}
    J(E) 
    &= \sigma_0 EP_{\rm LZ}(E) \nonumber \\
    &= \sigma_0 E e^{-E_c/|E|}. \label{eq:LZ_J} 
\end{align}
Here $\sigma_0=e^2v\tau/(\pi\hbar)$.

Eq.~\eqref{eq:LZ_J} is our first main result. For $|E|\ll E_c$, the current is exponentially small due to the singular factor $e^{-E_c/|E|}$, which is non-perturbative in $E$, and in particular, the linear response vanishes. The current turns on sharply near $|E|\sim E_c\propto\Delta^2$ due to the carrier generation through tunneling, and defining the onset by the peak of $\dv*[2]{J}{E}$ gives $E\approx0.336E_c$.

Eqs.~\eqref{eq:dist}, \eqref{eq:LZ_J} are valid when Eq.~\eqref{eq:E_cond} holds. In particular, the sharp onset around $|E|\sim E_c$ due to the carrier generation is described within the controlled approximation. This turn-on behavior is crucial for the discussion of the high harmonic transport below. 
We also note that Eqs.~\eqref{eq:dist}, \eqref{eq:LZ_J} are reasonably connected to the $E\to 0$ limit, where the scale separation fails.

Numerical solutions of Eq.~\eqref{eq:g_Boltzmann} confirm both Eq.~\eqref{eq:dist} and Eq.~\eqref{eq:LZ_J}, as shown in Fig.~\ref{fig:LZ_numerics_IV}: the conduction-band population follows the predicted exponential tail outside the tunneling region, and the resulting current quantitatively agrees with Eq.~\eqref{eq:LZ_J} over the entire field range, including $E\simeq 0$. 

Although related models have been studied by other methods~\cite{zener1934,landauQuantumMechanicsNonrelativistic1977,kaneZenerTunnelingSemiconductors1960,kitamuraCurrentResponseNonequilibrium2020}, to our knowledge Eq.~\eqref{eq:LZ_J} has not appeared previously. In Ref.~\cite{zener1934,landauQuantumMechanicsNonrelativistic1977, kaneZenerTunnelingSemiconductors1960}, the interband tunneling probability~\eqref{eq:LZ_prob} was derived for models of semiconductors, but they did not calculate the steady-state current, which requires the inclusion of relaxation effects. In Ref.~\cite{kitamuraCurrentResponseNonequilibrium2020}, the asymptotic expression for the steady-state current in general band insulators was derived with the nonequilibrium Green's function approach, but it did not show that the current is proportional to $EP_{\rm LZ}(E)$ for the model considered here (see Supplemental Material (SM) for consistency with Ref.~\cite{kitamuraCurrentResponseNonequilibrium2020}).

We note that the current--voltage relation approaches almost linear with a constant shift $|J|\simeq\sigma_0|E|-J_0$ at $|E|\gg E_c$, where $J_0=\sigma_0E_c$. In particular, when the system is gapless, the current response becomes completely linear. These are specific to one dimension and not the case in higher dimensions. On the other hand, the factor $e^{-E_c/|E|}$, and hence the sharp onset, persists in higher dimensions. We will discuss the case of higher dimensions later in this work. 

\begin{figure}
    \centering
    \includegraphics[width=\linewidth]{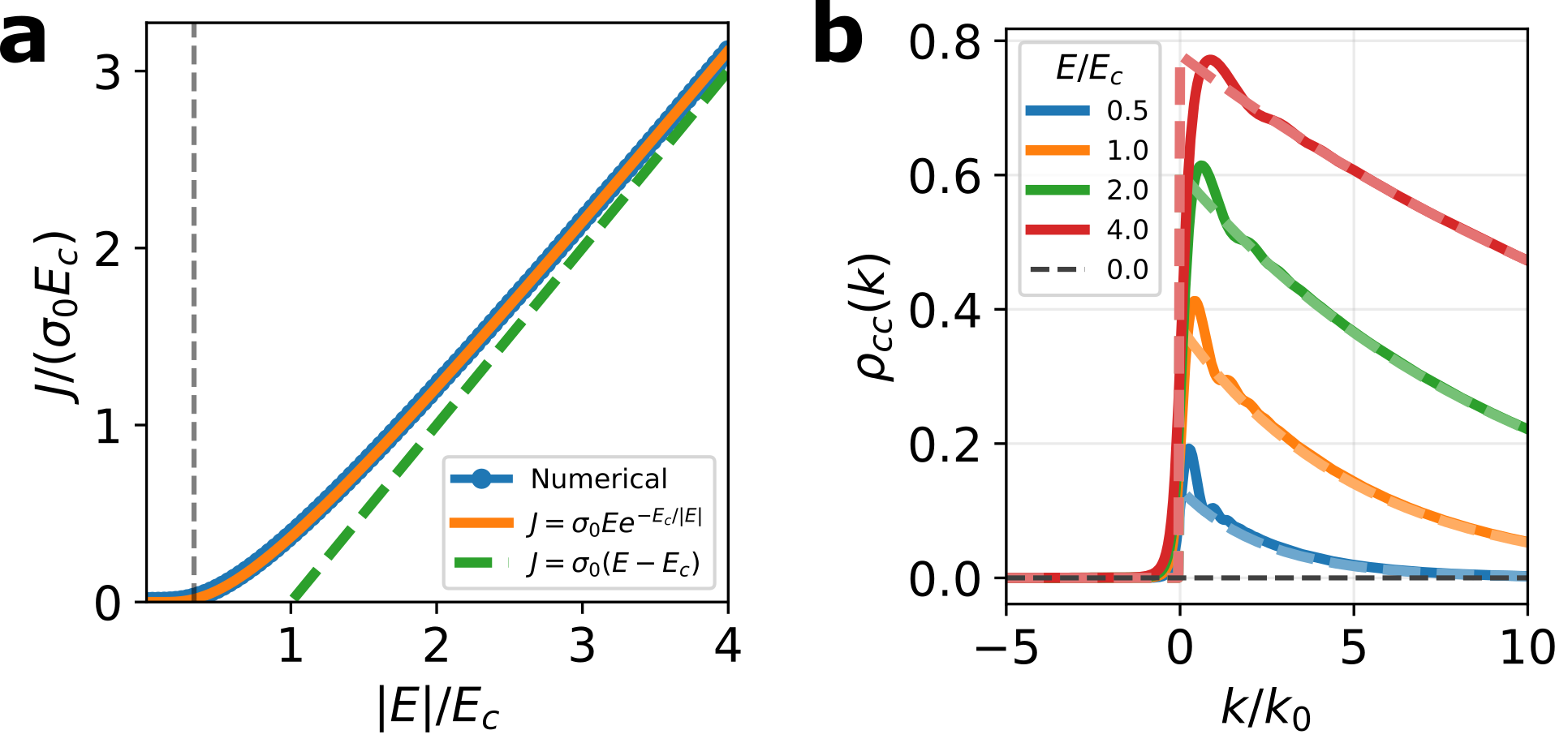}
    \caption{Comparison with the numerical solution of Eq.~\eqref{eq:g_Boltzmann2}. (a) Current--voltage relation of the 1D massive Dirac model. The orange line shows Eq.~\eqref{eq:LZ_J}; the black dashed line marks the onset field $E\approx0.336E_c$, defined by the peak of $\dv*[2]{J}{E}$. (b) Conduction-band population in momentum space. Dashed lines show Eq.~\eqref{eq:dist}. $k_0=\Delta/(\hbar v)$. Both calculations use $\tau=5.0\hbar/\Delta$.}
    \label{fig:LZ_numerics_IV}
\end{figure}

\textit{Comparison with diodes. ---} It is important to note that the singular current--voltage relation~\eqref{eq:LZ_J} is due to the carrier generation by interband tunneling.
This should be contrasted with other nonlinear current--voltage relations. 

For example, a PN diode exhibits the nonlinear and asymmetric response $J_{\rm PN}(E)=J_s(e^{E/E_T}-1)$ shown in Fig.~\ref{fig:HHG_PN_vs_LZ}(a), where $J_s$ is the saturation current density and $E_T$ is the threshold field~\cite{szePhysicsSemiconductorDevices2007}. $J_{\rm PN}$ arises from the diffusion and drift of the doped carriers across the junction, and the nonlinearity arises from the built-in potential within the junction. Since carriers are already present in equilibrium, $J_{\rm PN}$ differs fundamentally from the singular response~\eqref{eq:LZ_J} where the carriers are generated through tunneling under applied fields and no underlying inversion symmetry breaking is required. As a result, $J_{\rm PN}$ is non-singular at $E=0$ and, in particular, has a finite linear responses at small fields in contrast to Eq.~\eqref{eq:LZ_J}.

We note that a non-analytic current--voltage dependence $\exp(-a/|E|)$ also appears in a tunneling diode~\cite{kaneTheoryTunneling1961}. A tunneling diode discussed in Ref.~\cite{kaneTheoryTunneling1961} is made of a PN junction, and the carriers injected into the junction undergoes an interband tunneling, resulting in a nonlinear current--voltage relation with a singular factor $\exp(-a/|E|)$. 
However, it should be noted that our system is homogeneous and does not have any carriers in equilibrium in contrast to the tunneling diode; it rather generates carriers through interband tunneling under an external field. The singular response~\eqref{eq:LZ_J} is also symmetric, in contrast to the tunneling diode, which is in general asymmetric due to the inversion symmetry breaking.

So far, we have discussed DC transport due to the interband tunneling, and shown that it exhibits a sharp turn-on behavior characterized by a singular current--voltage relation. In the following, we discuss consequences of the singular current--voltage relation with a factor $\exp(-a/|E|)$ in AC responses. While we focus on the bulk systems with current--voltage relation~\eqref{eq:LZ_J}, qualitatively similar behavior is expected in tunneling diodes as well.

\textit{Singular high-harmonic transport and scaling law. ---}
We now discuss high-harmonic transport under an AC field $E(t)=E_0\cos(\omega t)$. In the quasi-static regime $\hbar\omega\ll\Delta$ and $\omega\tau\ll1$, the current follows the instantaneous DC relation, $J(t)=J(E(t))$. Its Fourier transform gives the harmonic amplitudes $J_n$. Because $J(E)$ is odd, only odd harmonics are nonzero. 

We show below that, the sharp turn-on behavior, characterized by the singular current--voltage relation, generate strong harmonics $J_n$ with large $n$ compared to non-singular current--voltage relation, and obey a scaling law that relates $J_n$ with different $n$. We call these behavior of high harmonics $J_n$ singular high-harmonic transport.

We first consider the $n$-dependence of $J_n$ at large $n$ for a fixed driving field $E_0$. We can analytically evaluate their asymptotic behavior with the saddle-point approximation. For the singular high-harmonic transport, the amplitude of harmonics $J_n$ can be written as
\begin{align}
    J_n &= \frac{2J_c E_0}{\pi E_c} \int_0^{\pi/2}\dd{\theta} e^{f(\theta)}, \label{eq:LZ_harmonic} \\
	f(z) &= inz + \log\cos z - \frac{E_c}{E_0\cos z}
\end{align}
where $J_c = \sigma_0 E_c$. According to the saddle point approximation~\cite{bender1999}, the integral can be approximated with the saddle point $z_*$ of $f(z)$, i.e., the solution of $\dv*{f}{z}=0$ on the complex plane. Choosing $z_*$ as the saddle point with the positive smallest $\Im z_*$, we can evaluate the integral at large $n$ as 
\begin{align}
	J_n \sim \frac{2J_c E_0}{\pi E_c} e^{f(z_*)} \label{eq:Jn_saddle}
\end{align}
up to a numerical factor. 

Let us find $z_*$. For a large $n\gg E_c/E_0$, $\dv*{f}{z}=0$ can be satisfied only when $\cos z$ is small, i.e., $z$ is close to $\pi/2$. At $z=\pi/2-x$, $\dv*{f}{z}$ behaves as 
\begin{align}
    \dv{f}{z} &= -\frac{E_c}{E_0x^2} - \frac{1}{x} + \qty(\frac{E_c}{6E_0} + i n)+\order{x}
\end{align}
Therefore, the solution of $\dv*{f}{z}=0$ at large $n$ is approximately given by $x_*\sim \sqrt{iE_c/(nE_0)}$ and correspondingly $z_*\sim \pi/2 - \sqrt{iE_c/(nE_0)}$. Plugging this into Eq.~\eqref{eq:Jn_saddle}, we find
\begin{align}
	J_n \sim \frac{2J_cE_0}{\pi E_c}\exp[-\sqrt{\frac{2nE_c}{E_0}}(1+i)] = \exp[-\order{\sqrt{n}}] \label{eq:LZ_Jn_asymptotic}
\end{align}

Eq.~\eqref{eq:LZ_Jn_asymptotic} shows that the singular high-harmonic transport decays only sub-exponentially at large $n \gg E_c/E_0$. The sharp onset behavior and the accompanying singularity in $J(E)$ is crucial to have strong high harmonics, because $J(E(t))$ must change rapidly in time to have a large $J_n$ at large $n$. 

The numerical calculation in Fig.~\ref{fig:HHG_PN_vs_LZ}(b) confirms this asymptotic behavior. For comparison, we show the harmonics from a PN diode, which has an analytic current--voltage relation. 
In contrast to the singular high-harmonic transport, the harmonics from a PN diode decay super-exponentially as $\exp[-\order{n\log n}]$, as the fitting in Fig.~\ref{fig:HHG_PN_vs_LZ}(b) shows.  
In general, we expect that harmonics from analytic current--voltage relations decay super-exponentially as $\exp[-\order{n\log n}]$ at large $n\gg E_0/E_T$, where $E_T$ is the characteristic electric field of the system (see SM for more details).

\begin{figure}
    \centering
    \includegraphics[width=0.9\columnwidth]{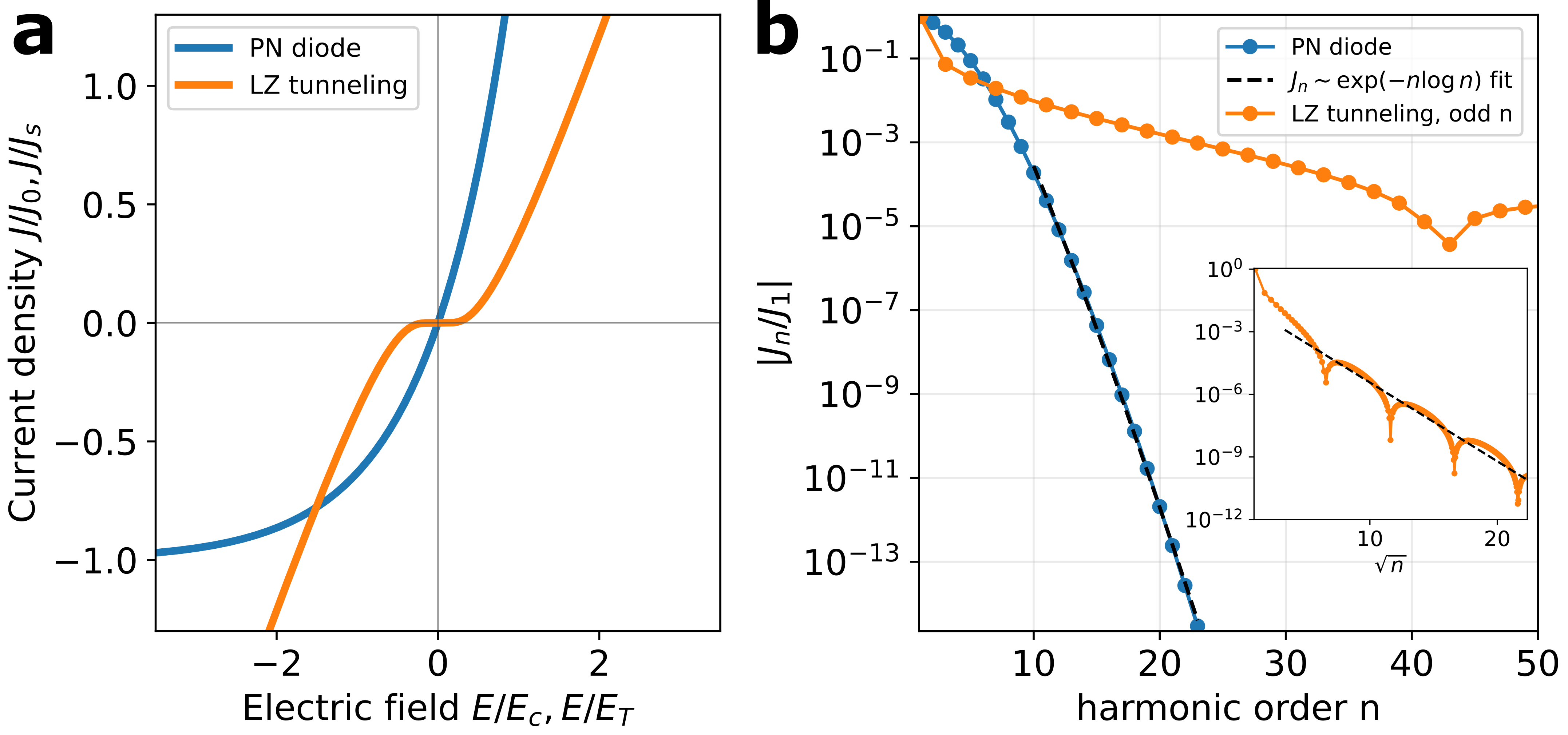} 
    \caption{Comparison of high-harmonic transport from 1D Landau-Zener (LZ) tunneling and an ideal PN diode. (a) Current--voltage relations for the PN diode, $J(E)=J_s[e^{E/E_T}-1]$, and LZ tunneling, Eq.~\eqref{eq:LZ_J}. (b) Harmonic amplitudes for $E_0=5.0$ and $E_T=E_c=1.0$. The dashed line shows the super-exponential decay $\exp[-\order{n\log n}]$; the dashed line in the inset shows the sub-exponential decay $\exp[-\order{\sqrt{n}}]$.}
	\label{fig:HHG_PN_vs_LZ}
\end{figure}

The same singularity further controls how $J_n$ depends on the driving field $E_0$ at a fixed large $n$.  
At large $n$, the contribution from the analytic current--voltage relation is negligible because it decays super-exponentially with $n$. Therefore, the high-harmonic amplitudes are determined solely by how fast the current is turned on around the singularity $|E(t)|\lesssim E_c$. Since only the contributions from where $\cos\omega t\sim 0$ matter when $E_0\gg E_c$, we can linearize $\cos\omega t$ near these points and then extend the integration range. This leads to (see SM for more details)
\begin{align}
    J_{n} 
    &\approx \frac{4(-1)^{\frac{n-1}{2}}}{n\pi}\int_0^{\infty}\dd{\theta}J(E_0\theta/n) e^{-\eta\theta}\sin\theta \label{eq:Jn_scaling_reasoning} 
\end{align}
$\eta=+0$ is the convergence factor. Eq.~\eqref{eq:Jn_scaling_reasoning} shows that $(-1)^{(n+1)/2}nJ_n$ depends on $E_0$ and $n$ only through $E_0/n$, leading to the following scaling law: 
\begin{align}
    \frac{J_n}{J_0} \approx \frac{(-1)^{\frac{n+1}{2}}}{n}f\qty(\frac{E_0}{nE_c}) \label{eq:scaling}
\end{align}
where $f(x)$ is independent of $n$. 

For the 1D current~\eqref{eq:LZ_J}, the scaling function can be obtained from Eq.~\eqref{eq:Jn_scaling_reasoning} as (see SM for the derivation)
\begin{align}
    f(x) = -\frac{4}{\pi}\Re K_2\qty(2\sqrt\frac{i}{x}) \label{eq:scaling_f}
\end{align}
where $K_\nu(z)$ is the modified Bessel function of the second kind. At $x\to0$ and $x\to \infty$, $f(x)$ for 1D current~\eqref{eq:scaling_f} asymptotically behaves as 
\begin{align}
    f(x\to 0) &= -\frac{2x^{1/4}}{\sqrt{\pi}}e^{-\sqrt{2/x}}\cos\qty(\sqrt{\frac{2}{x}}+\frac{\pi}{8}) \label{eq:1d_f0} \\
    f(x\to \infty) &= \frac{2}{\pi} \label{eq:1d_finf}
\end{align}

The scaling law naturally defines a crossover scale for electric field $E_0\sim nE_c \propto n\Delta^2$ at a fixed $n$ corresponding to $x\sim1$. 
For a fixed $n$, the large $E_0\gg n E_c \propto n\Delta^2$ limit corresponds to $f(x\to \infty)$ limit, and hence $J_n$ saturate to a constant. Note that saturation of $J_n$ at large $E_0$ is not the case in higher dimensions (see Table~\ref{tab:harmonic_scaling}). 

We note that the scaling law is consistent with the asymptotic behavior at large $n$ that we discussed earlier. For a fixed $E_0$, the large $n \gg n E_0/E_c$ limit corresponds to $f(x\to 0)$ limit, and hence $J_n$ decays sub-exponentially as $J_n\sim \exp[-\order{\sqrt{n}}]$, consistent with Eq.~\eqref{eq:LZ_Jn_asymptotic}.

The scaling law~\eqref{eq:scaling} and the asymptotic behavior~\eqref{eq:1d_f0}, \eqref{eq:1d_finf} are our second main result. They hold when $E_0 \gg E_c$, including both the large $E_0 (\gg nE_c)$ limit at a fixed $n$ and the large $n(\gg E_0/E_c)$ limit at fixed $E_0$. 
They clearly distinguish the singular high-harmonic transport from perturbative nonlinear transport~\cite{rikken2005,Pop2014,Sodemann2015,gao2014,ma2019a,gao2023,wang2023,Tokura2018,Ideue2017,lai2021,min2023a,chou2025a,orensteinTopologySymmetryQuantum2021,isobe2020,zhang2018a,Watanabe2020a}, where the $n$th harmonic scales as $J_n\propto E_0^n$ and cannot satisfy the scaling law and the asymptotic behavior.

The scaling law~\eqref{eq:scaling} as well as the scaling function~\eqref{eq:scaling_f} are confirmed in numerical calculations. Fig.~\ref{fig:HHG}(a) plots the odd harmonic amplitudes as functions of $E_0$, each with a sharp onset. Their onset shifts to larger fields and their magnitude decreases as $n$ increases. 
By rescaling the field and amplitude as $E_0/(nE_c)$ and $(-1)^{(n+1)/2}nJ_n/J_0$, respectively, the curves with different $n$ collapse as the scaling law~\eqref{eq:scaling} shows. The scaling function~\eqref{eq:scaling_f} is also plotted as a black dashed line in Fig.~\ref{fig:HHG}, which agrees well with $J_n$ at large $n$ and relatively large $E_0/(nE_c) \gtrsim 0.1$.

We note that the enhancement of the high harmonics and the scaling law originate from a current--voltage relation with a sharp onset behavior, and we expect that they do not depend on details of the singularity in the current--voltage relation (also see SM).

\begin{figure}
    \centering
    \includegraphics[width=0.8\columnwidth]{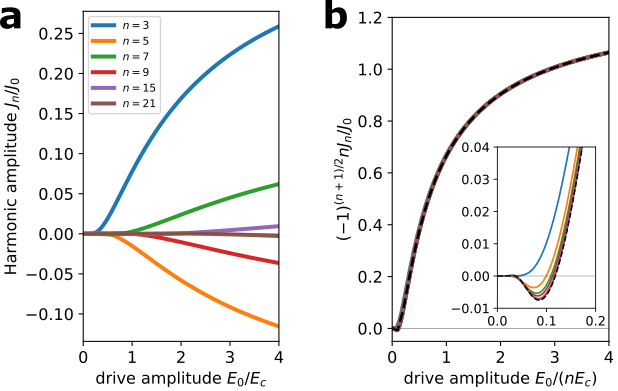}
    \caption{High-harmonic transport from 1D Landau-Zener tunneling. (a) Harmonic amplitude $J_n$ versus drive amplitude $E_0$. (b) Rescaled amplitude $(-1)^{(n+1)/2}nJ_n/J_0$ versus $E_0/(nE_c)$. The black dashed line is the scaling function $f(x)$ in Eq.~\eqref{eq:scaling_f}; the inset enlarges the small-$E_0/(nE_c)$ region.}
    \label{fig:HHG}
\end{figure}

\textit{Higher dimensions. ---}
Higher-dimensional systems can be treated as families of 1D momentum-space slices. In contrast to 1D cases, the singular response survives when the full system is gapless at an isolated point, because almost all slices retain a finite effective gap.

For example, near the gap minimum of a 2D system, the linearized $k\cdot p$ Hamiltonian is
\begin{align}
    H(\vec{k}) = \Delta \sigma_z + \hbar v(k_x\sigma_x + k_y\sigma_y).
\end{align}
Here we assumed an isotropic velocity for simplicity. After a unitary rotation of the Pauli matrices, each fixed-$k_y$ slice is a 1D massive Dirac Hamiltonian~\eqref{eq:1d_dirac} with effective mass $\Delta(k_y)=\sqrt{\Delta^2+\hbar^2v^2k_y^2}$. Therefore, the total current $J_x(E_x)$ can be obtained by integrating the 1D result~\eqref{eq:LZ_J} with the gap $\Delta(k_y)$ over $k_y$ as
\begin{align}
    J_x(E_x) 
    &= \frac{\sigma_0 k_0}{2\pi^{1/2}} E_x \sqrt{\frac{\abs{E_x}}{E_c}} e^{-E_c/|E_x|} \nonumber \\
    &\propto \mathrm{sgn}(E_x)|E_x|^{3/2}e^{-E_c/|E_x|}
\end{align}
where $k_0\equiv\Delta/(\hbar v)$. The field generates carriers in slices with $|k_y|\lesssim k_0\sqrt{|E_x|/E_c}$, producing the extra factor $\sqrt{|E_x|/E_c}$. In the limit $\Delta\to0$, the result reduces to the known graphene scaling $J_x\propto\operatorname{sgn}(E_x)|E_x|^{3/2}$~\cite{allorSchwingerMechanismGraphene2008,vandecasteeleCurrentvoltageCharacteristicsGraphene2010,berdyuginOutofequilibriumCriticalitiesGraphene2022}.

More generally, the longitudinal current of a linearized $d$-dimensional $k\vdot p$ model is
\begin{align}
    &J_x(E_x) \propto \mathrm{sgn}(E_x)|E_x|^{(d+1)/2} e^{-E_c/|E_x|}, \quad E_c\propto \Delta^2, \label{eq:Jx_general_d}
\end{align}
For $d>1$, the gapless limit remains nonanalytic and nonlinear, unlike the 1D result. For $d=3$, a similar tunneling-generated carrier density was derived previously for semiconductors~\cite{kaneZenerTunnelingSemiconductors1960}.

\begin{table}[t]
\centering
\begin{tabular}{@{}l @{\hspace{1.2em}} l c c@{}}
\toprule
$d$
&
$J(E)$
&
$\abs{J_n\!\left(\frac{E_0}{nE_c}\to 0\right)}$
&
$\abs{J_n\!\left(\frac{E_0}{nE_c}\to \infty\right)}$
\\
\midrule
$1$
&
$E e^{-E_c/|E|}$
&
$\exp\!\left[-\order{\sqrt{\frac{nE_c}{E_0}}}\right]$
&
$ \frac{1}{n}\times \mathrm{const.}$
\\
$2$
&
$E^{3/2} e^{-E_c/|E|}$
&
$\exp\!\left[-\order{\sqrt{\frac{nE_c}{E_0}}}\right]$
&
$\frac{1}{n}\left(\frac{E_0}{nE_c}\right)^{3/2}$
\\
$3$
&
$E^{2} e^{-E_c/|E|}$
&
$\exp\!\left[-\order{\sqrt{\frac{nE_c}{E_0}}}\right]$
&
$\frac{1}{n}\left(\frac{E_0}{nE_c}\right)^{2}$
\\
\bottomrule
\end{tabular}
\caption{Asymptotic scaling of the harmonic amplitude $J_n$ for linear $k\cdot p$ models in spatial dimension $d$. See SM for the derivation.}
\label{tab:harmonic_scaling}
\end{table}

The asymptotic harmonics obtained from Eq.~\eqref{eq:Jx_general_d} are summarized in Table~\ref{tab:harmonic_scaling}. For gapped systems, the sub-exponential decay $\exp{-\order{\sqrt{n}}}$ and the scaling law~\eqref{eq:scaling} persist in any dimension, with a dimension-dependent scaling function $f$ given in SM.

\textit{Discussion. ---} 
Singular high-harmonic transport differs from optical high-harmonic generation at optical frequencies~\cite{goulielmakisHighHarmonicGeneration2022,tamayaDiabaticMechanismsHigherOrder2016,yoshikawaHighharmonicGenerationGraphene2017}: singular high-harmonic transport operates in the quasi-static regime $\omega\tau\ll1$, typically below the THz scale, and is frequency-independent within that regime. It also requires much weaker fields than optical pulses, whose peak amplitudes are often $\sim\SI{1e10}{V/m}$. 

We expect singular high-harmonic transport to arise generically in small-gap or gapless systems, including Dirac/Weyl semimetals, topological-insulator surface states, and systems near topological phase transitions. Two-dimensional materials with tunable gaps are especially promising.
For a gap of $\Delta\sim\SI{1}{meV}$ and the velocity $v\sim \SI{1e6}{m/s}$, $E_c\sim\SI{5}{kV/m}$, which is accessible in standard transport experiments.

We also note that a recent bilayer-graphene experiment observed Hall-current harmonics up to the 300th order at finite displacement field and carrier density~\cite{zan2024}. Although our theory treats longitudinal response without equilibrium carriers, the observed huge nonlinearity may have a Landau-Zener tunneling origin.

For large gap systems, the required field is large and may activate competing mechanisms. For example, an energy gap of $\Delta\gtrsim\SI{0.1}{\electronvolt}$ and velocity $v\sim\SI{1e6}{m/s}$ give $E_c\sim\SI{50}{MV/m}$, where avalanche breakdown may occur~\cite{sun2012}. When $\Delta$ is comparable to optical-phonon energies, $\sim\SI{10}{meV}$--$\SI{100}{meV}$, inelastic phonon scattering may also require explicit treatment~\cite{liNonequilibriumExcitationsTransport2018}.

Throughout this work, we assume zero temperature. At finite temperature, we expect that the thermally excited carriers would round off the current--voltage relation and the enhancement in high harmonics and the scaling law would be cut off at large $n$. Our results discussed here are valid for sufficiently low temperatures relative to the system's energy gap $\Delta$.

\begin{acknowledgements}
This work was supported by National Science Foundation (NSF) Convergence Accelerator Award No. 2235945.
\end{acknowledgements}

\bibliography{Extracted_ref}

\begin{widetext}

\clearpage
\onecolumngrid %

\setcounter{secnumdepth}{3}
\setcounter{tocdepth}{3}
\setcounter{section}{0}
\setcounter{subsection}{0}
\setcounter{subsubsection}{0}

\renewcommand{\thesection}{S\arabic{section}}
\renewcommand{\thesubsection}{\thesection.\arabic{subsection}}

\renewcommand{\thesubsubsection}{\thesubsection.\arabic{subsubsection}}

\setcounter{equation}{0}
\setcounter{figure}{0}
\setcounter{table}{0}

\renewcommand{\theequation}{S\arabic{equation}}
\renewcommand{\thefigure}{S\arabic{figure}}
\renewcommand{\thetable}{S\arabic{table}}

\makeatletter
\@ifpackageloaded{hyperref}{%
  \renewcommand{\theHsection}{SM.\arabic{section}}
  \renewcommand{\theHsubsection}{\theHsection.\arabic{subsection}}
  \renewcommand{\theHequation}{SM.\arabic{equation}}
  \renewcommand{\theHfigure}{SM.\arabic{figure}}
  \renewcommand{\theHtable}{SM.\arabic{table}}
}{}
\makeatother

\begin{center}
  \large\bfseries
  Supplemental Material for ``Singular high-harmonic transport above a quantum threshold''
\end{center}

\makeatletter
\section*{Supplemental contents}
\begingroup
  \def\l@section{\@dottedtocline{1}{0em}{3em}}
  \def\l@subsection{\@dottedtocline{2}{1.5em}{4em}}
  \def\l@subsubsection{\@dottedtocline{3}{3em}{5em}}
  \@starttoc{smtoc}
\endgroup

\let\SMoriginaladdcontentsline\addcontentsline
\renewcommand{\addcontentsline}[3]{%
  \def\SMext{#1}\def\SMtoc{toc}%
  \ifx\SMext\SMtoc
    \SMoriginaladdcontentsline{smtoc}{#2}{#3}%
  \else
    \SMoriginaladdcontentsline{#1}{#2}{#3}%
  \fi
}
\makeatother

\title{Supplemental Materials for ``Singular high harmonic transport above quantum threshold''}
\date{\today}

\maketitle

\section{Consistency with Ref.~\cite{kitamuraCurrentResponseNonequilibrium2020}}
In Ref.~\cite{kitamuraCurrentResponseNonequilibrium2020}, the nonperturabative current in band insulators is calculated with nonequilibrium Green's function approach. At zero temperature, it was shown that the current is given by 
\begin{align}
	J_{\rm LZ} &= \mp P_{\rm LZ}\int_{\mp \Lambda}^0 \frac{\dd{k}}{2\pi} (\partial_k \Delta) e^{\hbar k/(eE\tau)} \label{eq:kitamura_J}
\end{align}
Here, we replace the dissipation $\Gamma$ in Ref.~\cite{kitamuraCurrentResponseNonequilibrium2020} with $\tau/2$. We consider only the intraband contribution, since we assume a long relaxation time $\tau\Delta/\hbar \gg 1$ and the interband contribution calculated in Ref.~\cite{kitamuraCurrentResponseNonequilibrium2020} is small, as pointed out in Ref.~\cite{kitamuraCurrentResponseNonequilibrium2020}. 

For our model with large $\Delta \tau/\hbar$ in the main text, the dominant contribution in Eq.~\eqref{eq:kitamura_J} comes from large $k$ where $\Delta \approx \hbar v k$. Then, we obtain the same result as ours from Eq.~\eqref{eq:kitamura_J}. Therefore, our results are consistent with Ref.~\cite{kitamuraCurrentResponseNonequilibrium2020}.

We note that, Ref.~\cite{kitamuraCurrentResponseNonequilibrium2020} claims that the (intraband) tunneling current in the Landau-Zener model is propotional to $E^2 P_{\rm LZ}$, rather than $E P_{\rm LZ}$. However, this claim is based on further approximation of Eq.~\eqref{eq:kitamura_J} to obtain a series in $eE\tau/\hbar$ and does not apply to our case. 

\section{Remark on Quantum Master Equation}
\newcommand{\rhoS}{\tilde{\rho}}
Here, we present several remarks on the quantum master equation we used in the main text.

\subsection{Relation to previous literature}
In the main text, we used the following equation for the one-body reduced density matrix $\rho(k,t)$:
\begin{align}
    \pdv{\rho}{t} + \dot{k}\pdv{\rho}{k} = -\frac{i}{\hbar}\comm{H(k)}{\rho} - \frac{\rho-\rho_0(k)}{\tau}. \label{ap:eq:master}
\end{align}
where $\rho_0(k)$ is the reduced density matrix in equilibrium, and $\tau$ is the relaxation time. 

Eq.~\eqref{ap:eq:master} may look different from the expressions found in the literature. For example, in Ref.~\cite{satoLightinducedAnomalousHall2019}, the following equation is used:
\begin{align}
    \dv{}{t}\rhoS(K(t), t) &= -\frac{i}{\hbar}\comm{H(K(t))}{\rhoS(K(t),t)} - \frac{\rhoS(K(t),t)-\rho_{0}(K(t))}{\tau} \label{ap:eq:master_S}\\
    K(t) &= k-\frac{e}{\hbar}A(t)
\end{align}
where we change the notations from Ref.~\cite{satoLightinducedAnomalousHall2019} to align with ours, and we denote the reduced density matrix with $\rhoS$ to distinguish the one for our master equation~\eqref{ap:eq:master}. 

However, Eq.~\eqref{ap:eq:master_S} is equivalent to Eq.~\eqref{ap:eq:master} by identifying $\rhoS$ with $\rho$ as 
\begin{align}
    \rhoS(K(t), t) = \rho(k=K(t), t)
\end{align}
Then the time derivative of $\rhoS(K(t), t)$ can be written as 
\begin{align}
    \dv{}{t}\rhoS(K(t), t) = \pdv{\rho}{t} + \dot{k} \pdv{\rho}{k}
\end{align}
Therefore, Eq.~\eqref{ap:eq:master} follows from Eq.~\eqref{ap:eq:master_S}.

\subsection{Applicability of the relaxation time approximation}
In ref.~\cite{teradaLimitationsImprovementsRelaxation2024}, it was pointed out that the relaxation time approximation can produce a finite linear response, which is unphysical. However, the produced finite current is of order $(\tau \Delta)^{-1}$, and we can safely apply the relaxation time approximation when the band gap $\Delta$ is well defined, i.e., $\tau \gg \hbar/\Delta$, as we assumed in the main text.

\section{High harmonics from analytic current-voltage relations}
Here, we provide a detailed analysis of high harmonics from analytic current-voltage relations. In particular, we show that it typically decays with $n$ super-exponentially as:
\begin{align}
    J_n \sim \exp(-\order{n\log n}).
\end{align}

Suppose a current-voltage relation $J(E)$ is analytic so that we can expand it around $E=0$:
\begin{align}
    J(E) &= \sum_{n=1}^{\infty} \frac{1}{n!} J^{(n)} E^n
\end{align}
where $J^{(n)}=\eval{\pdv*[n]{J}{E}}_{E=0}$. When an AC electric field $E(t)= E_0 \cos\omega t$ is applied, the current $J(t)$ is given by 
\begin{align}
    J(t) &= \sum_{n=1}^{\infty} \frac{E_0^n}{n!} J^{(n)}  \cos^n\omega t \nonumber \\
    &= \sum_{n=1}^{\infty} \frac{E_0^n}{2^nn!} J^{(n)} \sum_{k=0}^n \begin{pmatrix}
        n \\
        k
    \end{pmatrix} e^{i(n-2k)\omega t} \label{eq:Jt_taylor}
\end{align}

We define $n$th harmonic of $J(t)$ as 
\begin{align}
    J_n &= \int_0^{2\pi/\omega}\frac{\dd{t}}{2\pi/\omega} J(t) e^{in\omega t}
\end{align}
From Eq.~\eqref{eq:Jt_taylor}, we find 
\begin{align}
    J_n 
    &= \sum_{m=n, m+n:\mathrm{even}}^{\infty} \frac{E_0^m}{2^mm!} J^{(m)} \begin{pmatrix}
        m \\
        (n+m)/2
    \end{pmatrix} \\
    &= \sum_{k=0}^{\infty}  \frac{E_0^{n+2k}}{2^{n+2k}(n+2k)!} J^{(n+2k)} \begin{pmatrix}
        n+2k \\
        n+k
    \end{pmatrix} \\
    &= \frac{E_0^n}{2^n}\sum_{k=0}^{\infty}  \frac{E_0^{2k}}{2^{2k}(n+k)!k!} J^{(n+2k)} 
\end{align}

Denoting the characteristic amplitude of the current and the electric field of the system as $J_0$ and $E_T$, $J^{(n)}$ can be estimated as $J^{(n)}\sim J_0/E_T^n$. Therefore, 
\begin{align}
    J_n 
    &\sim J_0\qty(\frac{E_0}{2E_T})^n\sum_{k=0}^{\infty}  \frac{1}{(n+k)!k!} \qty(\frac{E_0}{2E_T})^{2k} \label{eq:J_n_taylor}
\end{align}

At large $n$, the summation in Eq.~\eqref{eq:J_n_taylor} is dominated by $k=0$ term, because each term in the summation is bounded as 
\begin{align}
    \abs{\frac{1}{(n+k)!k!} \qty(\frac{E_0}{2E_T})^{2k}} \le \frac{1}{n! n^k k!}\qty(\frac{E_0}{2E_T})^{2k} = \frac{1}{n! k!}\qty(\frac{E_0}{2nE_T})^{2k}
\end{align}
Therefore, at large $n$ satisfying $n\gg E_0/E_T$, terms with $k>0$ in the summation are suppressed and the sum is dominated by the $k=0$ term. Hence 
\begin{align}
    J_n &\sim \frac{J_0}{n!}\qty(\frac{E_0}{2E_T})^n \sim J_0\exp[-n\log n + n(1+\log(\frac{E_0}{2E_T}))] = \exp(-\order{n\log n}) \label{eq:J_n_taylor_k0}
\end{align}
This is the result used in the main text.

\section{Saddle-point approximation analysis for asymptotics of high harmonics}
In this section, we present a detailed derivation of the asymptotic expressions of high harmonics based on the saddle-point approximation given in the main text. We discuss both PN diodes and 1D singular high-harmonic transport. 

\subsection{PN diode}
Consider an ideal PN junction. Its current-voltage relation is given by 
\begin{align}
    J(E) = J_s (e^{E/E_T}-1)
\end{align}

When an AC field $E(t)=E_0\cos \omega t$ is applied, the current $J(t)=J(E(t))$ contains many harmonics. Its $n$th harmonic is given by
\begin{align}
    J_n &= \int_0^{2\pi/\omega}\frac{\dd{t}}{2\pi/\omega} e^{in\omega t} J(E_0\cos\omega t) \\
    &= \frac{J_s}{2\pi} \int_0^{2\pi}\dd{\theta} \exp[in\theta + \frac{E_0}{E_T}\cos\theta]
\end{align}

To analyze the harmonic, we employ a saddle point approximation. Noting that the integrand is analytic with respect to $\theta$ in the complex plane (except the point at infinity), we can analytically continue $\theta\to z\in \mathbb{C}$, and we can deform the integration path. Then we can approximate the integral by a saddle point of the argument in the exponential, and we find 
\begin{align}
    &J_n \approx \frac{J_s}{2\pi} \exp[-n(x_0 - \frac{E_0}{nE_T}\cosh x_0)] \\
    &\sinh x_0 = \frac{nE_T}{E_0}
\end{align}

At small $E_0\ll nE_T$, $J_n$ can be further approximated as 
\begin{align}
    &J_n(E_0\ll nE_T) \nonumber \\
    &\approx \frac{J_s}{2\pi} \exp[-n\log\frac{2nE_T}{E_0} - n - \frac{E_0^2}{2nE_T^2}+\order{n^{-2}(E_0/E_T)^3}] \nonumber \\
    &\propto  \frac{J_s}{2\pi} \qty(\frac{E_0}{2nE_T})^{n}
\end{align}

At large $E_0\gg nE_T$, $J_n$ reduces to 
\begin{align}
    J_n(E_0\gg nE_T) &\approx \frac{J_s}{2\pi} \exp[\frac{E_0}{E_T} - \frac{n^2E_T}{2E_0}]
\end{align}

Either $E_0\gg n E_T$ or $E_0 \ll nE_T$, $J_n$ decays faster than exponential in $n$. In the large $n$ limit, $E_0 \ll n E_T$ and thus $J_n \sim n^{-n}$.

\subsection{1D Singular high harmonic transport}
Consider 1D massive Dirac Hamiltonian and its transport through LZ tunneling:
\begin{align}
    J(E) = \sigma_0 E \exp(-E_c/|E|)
\end{align}
When an AC voltage $E(t)=E_0\cos\omega t$ is applied, the $n$th harmonic is given by 
\begin{align}
    J_n &= \frac{2J_c E_0}{\pi E_c} \int_0^{\pi/2}\dd{\theta} \exp[in\theta + \log\cos\theta - \frac{E_c}{E_0|\cos\theta|}] \label{ap:eq:LZ_harmonic}
\end{align}
where $J_c = \sigma_0 E_c$.

The contribution around maximum of $|E(t)|$ gives $\sim \exp(-\mathrm{const.}\times n^2)$ at large $n$. There is a more important contribution from the nonanalyticity, $E(t)\sim 0$. 

First, let us consider the contribution around maximum $|E(t)|$, which corresponds to $\theta\sim 0$. Around $\theta\sim0$, we can expand $\cos\theta=1-\theta^2/2$. Plugging this into Eq.~\eqref{ap:eq:LZ_harmonic} and applying the saddle approximation, we obtain
\begin{align}
    J_n \sim \frac{J_c E_0}{\pi E_c} \exp[-\frac{n^2}{2(1+(E_0/E_c)^{-1})} - \frac{E_c}{E_0}]
\end{align}

Now let us consider the contribution around $E(t)\sim0$, which corresponds to $\theta\sim\pi/2$. For $\theta=\pi/2-x$, we can expand $\cos\theta=x$, and the integral can be rewritten as 
\begin{align}
    J_n &= \frac{2J_c E_0e^{in\pi/2}}{\pi E_c} \int_0^{\pi/2}\dd{x} \exp[-in x + \log x - \frac{E_c}{E_0|x|}]
\end{align}

At large $n$, the saddle point with the smallest $|x|$ is approximately given by $x\approx \sqrt{iE_c/(nE_0)}$. Then the saddle point approximation yields the asymptotics at large $n$ (the strongest dependence on $n$) as
\begin{align}
    J_n \sim \frac{2J_cE_0}{\pi E_c}\exp[-\sqrt{\frac{nE_c}{2E_0}}(1+i)]
\end{align}
Therefore, the contribution due to the singularity at $\theta=\pi/2$ or $E(t)\sim 0$ is more dominant than the maximum $|E(t)|$.

\section{Derivation of scaling law for singular high-harmonic transport}
Here we present the detailed derivation of the scaling law for singular high-harmonic transport discussed in the main text. 

We start with the definition of the harmonics, given by the Fourier transform of the current $J(E(t))$:
\begin{align}
    J_{n} 
    &= \int_0^{2\pi/\omega}\frac{\dd{t}}{\pi/\omega} J(E_0\cos\omega t) \cos (n\omega t) 
\end{align}
By changing the variable from $t$ to $\theta=\omega t$, we find
\begin{align}
    J_{n} 
    &= \int_0^{2\pi}\frac{\dd{\theta}}{\pi} J(E_0\cos\theta) \cos (n\theta)
\end{align}
Noting that the singular response from the interband tunneling is symmetric, $J(E)=-J(-E)$, we can rewrite $J_n$ as 
\begin{align}
    J_n &= \frac{4(-1)^{\frac{n-1}{2}}}{\pi}\int_0^{\pi/2}\dd{\theta} J(E_0\sin\theta) \sin (n\theta) \label{ap:eq:fourier}
\end{align}

Now, we note that the main contribution to $J_n$ comes from around the singularity of $J(E)$. As shown in the main text, the analytic current--voltage relation decays fast at high-order harmonics. The rapid oscillations of $\sin(n\theta)$ cancel contributions from smooth portions of $J(t)$, leaving the least-smooth region due to the singularity near the zero crossings of $E_0\sin\theta$. 

This observation motivates us to approximate the integral with the expansion $\sin\theta\simeq \theta$, and then extend the integration range from $[0, \pi/2]$ to $[0,\infty]$. These approximations lead to the following approximation:
\begin{align}
    J_n &\approx \frac{4(-1)^{\frac{n-1}{2}}}{n\pi}\int_0^{\infty}\dd{\theta}J(E_0\theta/n) e^{-\eta\theta}\sin\theta \label{ap:eq:Jn_scaling_reasoning}
\end{align}

Eq.~\eqref{ap:eq:Jn_scaling_reasoning} shows the relation:
\begin{align}
    J_n = \frac{(-1)^{\frac{n-1}{2}}}{n} f\qty(\frac{E_0}{nE_c})
\end{align}
This is the scaling law discussed in the main text. 

We emphasize that this result holds for a general current--voltage relation that has a singularity around $E=0$. More specifically, when the electric field at which the singularity is located, $E_{\rm sing}$, is much smaller than $E_0$, the scaling law derived above holds.

\section{Examples of singular high-harmonic transport}
\subsection{Singular high-harmonic transport from interband tunneling in general dimensions $d$}
\subsubsection{Scaling law and Scaling function for Singular high-harmonic transport}
Here we show that the singular high-harmonic transport from the interband tunneling in general dimension $d$ follows the scaling law given by
\begin{align}
    J_n(E_0)/J_0 &= \frac{(-1)^{(n+1)/2}}{n} f\qty(\frac{E_0}{nE_c}) \label{ap:eq:scaling_law}\\ 
    f(x) &= -\frac{4}{\pi} \Re\qty[(-ix)^{(d-1)/4}K_{(d+3)/2}\qty(2\sqrt{\frac{i}{x}})] \label{ap:eq:scaling_func}
\end{align}

Generally, the current-voltage relation for general dimensions $d$ of the singular high-harmonic transport is 
\begin{align}
    J_d(E) = \sigma_{d,0}E\abs{\frac{E}{E_c}}^{(d-1)/2} e^{-E_c/|E|}
\end{align}
where $\sigma_{d,0}$ has a unit of $d$-dimensional conductivity, or in a dimensionless form,
    $J_d/J_0 = \frac{E}{E_c}\abs{\frac{E}{E_c}}^{(d-1)/2} e^{-E_c/|E|}$,
with $J_0=\sigma_{d,0}E_c$.

When $E(t)=E_0\cos\omega t$ is applied, $n$th harmonic of the current is finite only for odd $n$ and always real, given by 
\begin{align}
    J_n/J_0 &= \Re\int_0^{2\pi/\omega}\frac{\dd{t}}{2\pi/\omega} J(E_0\cos\omega t)e^{in\omega t} \nonumber \\
    &= \Re\qty[\frac{2i^{n}}{\pi}\qty(\frac{E_0}{E_c})^{(d+1)/2} \int_0^{\pi/2}\dd{x}e^{-inx}\sin^{(d+1)/2}x e^{-E_c/(E_0\sin x)}].
\end{align}
Since the singularity is around $x=0$ of the integrand, we approximate $\sin x\approx x$ and extend the integral range from $[0, \pi/2]$ to $[0,\infty]$. Further changing the variable as $x'=E_0x$, we obtain
\begin{align}
    J_n/J_0 &\approx \Re\qty(\frac{2 i^{n}E_c}{\pi E_0}\int_0^{\infty}\dd{x}x^{(d+1)/2} \exp[-\frac{1}{x}-\frac{inE_c}{E_0}x])
\end{align}
This integral can be rewritten with the modified Bessel function of the second kind, $K_{\nu}(z)$. Useful formulas and properties are summarized in Sec.~\ref{ap:subsec:Knu}. 
Comparing a formula~\eqref{ap:eq:Knu} with the expression for $J_n$, we find
\begin{align}
    J_n/J_0 \approx \Re\qty[\frac{4 i^{n}E_c}{\pi E_0} \gamma_{\eta}^{-\nu/2} K_{\nu}(2\sqrt{\gamma_{\eta}})]
\end{align}
with $\nu=(d+3)/2$ and $\gamma_{\eta} = \eta + inE_c/E_0$. $\eta\to 0+$ is a convergence factor. From this form, we confirm the scaling law~\eqref{ap:eq:scaling_law} and the scaling function~\eqref{ap:eq:scaling_func}.

\subsubsection{Asymptotic behavior of the scaling function}
The asymptotic behavior of $f(x)$ can be found using the formulas in Sec.~\ref{ap:subsec:Knu}. For $x \to 0$, we find 
\begin{align}
    f(x\to 0) &= -\frac{2}{\sqrt{\pi}}\Re\qty[(-ix)^{d/4}e^{-2\sqrt{i/x}}]
\end{align}
and for $x\to \infty$, we find 
\begin{align}
    f(x\to \infty) &= \begin{cases}
        \frac{2}{\pi} & (d=1) \\
        \frac{-2}{\pi}\Gamma\qty(\frac{d+3}{2})\cos(\frac{d+1}{4}\pi) x^{\frac{d+1}{2}} & (d>1)
    \end{cases}
\end{align}
Note that, to derive the expression for $d=1$, we need the next leading order of $K_2(z)$ at small $z$, given as Eq.~\eqref{ap:eq:K2}.

\subsection{Current--voltage relation with sharp onset}
In the main text, we discuss singular high-harmonic transport due to the sharp onset behavior of the current due to the carrier generation through interband tunneling.

Here, we discuss another example, where the current--voltage relation is given by 
\begin{align}
    J(E) &= \mathrm{sgn}\sigma_0 \max(|E|-E_c, 0).
\end{align}
The current is zero for $|E|\le E_c$, and turns on at $|E|=E_c$ linearly. This current--voltage relation can be viewed as a crude approximation of the current--voltage relation due to the tunneling in one dimension derived in the main text.

For this current--voltage relation, we can readily calculate the harmonic amplitude under a drive field $E(t)=E_0\cos\omega t$. The $n$th harmonic $J_n$ for $E_0>E_c$ is then given by
\begin{align}
    J_n &= \frac{4\sigma_0}{\pi}\qty[\frac{E_0}{2}\qty(\frac{\sin[(n-1)\alpha]}{n-1} + \frac{\sin[(n+1)\alpha]}{n+1})-\frac{E_c}{n}\sin(n\alpha)]
\end{align}
where $\alpha = \arccos(E_c/E_0)$. In particular, at large $n\gg 1$, we can expand in $1/n$ as
\begin{align}
    J_n &\simeq \frac{4\sigma_0}{\pi}\qty[\frac{E_0}{2}\qty(\frac{\sin[(n-1)\alpha]}{n} + \frac{\sin[(n+1)\alpha]}{n})-\frac{E_c}{n}\sin(n\alpha)] \nonumber \\
    &+ \frac{4\sigma_0}{\pi}\qty[\frac{E_0}{2}\qty(\frac{\sin[(n-1)\alpha]}{n^2} - \frac{\sin[(n+1)\alpha]}{n^2})] + \order{1/n^3}
\end{align}
We can show that the first line vanishes and thus the leading order in $1/n$ is $\order{1/n^2}$. Therefore, we find 
\begin{align}
    J_n &\simeq -\frac{4\sigma_0 E_0}{\pi n^2}\cos(n\alpha)\sin\alpha + \order{1/n^3}
\end{align}
Therefore, apart from the oscillating component due to $\cos(n\alpha)$, $J_n$ decays with $n$ as $1/n^2$, and thus the high harmonic transport is significantly enhanced compared to an analytic current--voltage relation.

We also note that, at large $E_0/E_c\gg 1$, since $\alpha\approx \pi/2 - E_c/E_0$, we find 
\begin{align}
    J_n &\approx (-1)^{(n+1)/2}\frac{4\sigma_0 E_0}{\pi n^2} \sin(n E_c/E_0) =  \frac{(-1)^{(n+1)/2}}{n}f\qty(\frac{E_0}{nE_c}) \\
    f(x) &= \frac{4\sigma_0 E_c}{\pi} x \sin(1/x) 
\end{align}
Therefore, at large $E_0$, $J_n$ follows the same scaling law as shown in the main text.  

\section{Mathematical properties of functions appeared in the main text}
In this section, we list useful facts about the functions that appear in the main text. 
\subsection{Derivatives of $|z|^{\alpha}e^{-1/|z|}$}
As shown in the main text, the Landau-Zener tunneling results in a current-voltage relation in the following functional form:
\begin{align}
    f(z) = |z|^{\alpha}e^{-1/|z|} 
\end{align}

Here we show that $n$th derivative of $f(z)$ is given by
\begin{align}
    \dv[n]{f}{z} = (-\mathrm{sgn}(z))^n  n! e^{-1/|z|} |z|^{\alpha-n} L_n^{-\alpha-1}(1/|z|)
\end{align}
where $L_n^{\alpha}(z)$ is the associated Laguerre polynomial. For $\alpha=0$, 

\textit{Proof.} --- We focus on $z>0$. The derivative for $z<0$ can be easily obtained by noting that $f(z)$ is an even function. 

We set $u=1/z$. Then $f(z)=u^{-\alpha}e^{-u}$ and $\dv*{x}=-u^2\dv*{u}$. A convenient identity is 
\begin{align}
    \dv[n]{z} g(1/z) = (-1)^n u^{n+1}\dv[n]{u}[u^{n-1} g(u)],
\end{align}
which can be proved by induction. Setting $g(u)=u^{-\alpha}e^{-u}$ gives 
\begin{align}
    \dv[n]{x} (z^{\alpha}e^{-1/z}) = (-1)^n u^{n+1}\dv[n]{u}[u^{n-1-\alpha} e^{-u}].
\end{align}

Now we use the Rodrigues formula for the associated Laguerre polynomials $L_n^\alpha$~\cite{NIST:DLMF}:
\begin{align}
    L_n^{\beta}(u) = \frac{u^{-\beta}e^u}{n!}\dv[n]{u}[u^{n+\beta}e^{-u}].
\end{align}
Choosing $\beta = -1-\alpha$, we find 
\begin{align}
    \dv[n]{x} (z^{\alpha}e^{-1/z}) = (-1)^n z^{\alpha-n}e^{-1/z} n! L_n^{-\alpha-1}(1/z). 
\end{align}
The derivative for $z<0$ is given by multiplying this by $\mathrm{sgn}(z)^{n}$. \qedsymbol

\subsection{High-order derivatives of the current-voltage relation for LZ tunneling}
The large high-harmonic transport from the current-voltage relation for the tunneling,
\begin{align}
    J(E) = \sigma_0 E \exp(-E_c/|E|),
\end{align}
is hinted in its high-order derivatives. Even though it is continuous, the sharp turn-on behavior around $E\sim E_c$ results in a sharp peak of higher-order derivatives of $J(E)$. Since $n$th order derivative of $P_{\rm LZ}$ at small $E/E_c$ scales as 
    $\dv*[n]{P_{\rm LZ}}{E} \sim \order{E^{-2n}e^{-E_c/|E|}}$, and thus
the $n$th derivative of $J(E)$ behaves as:  
\begin{align}
    \dv[n]{J}{E} \sim \frac{1}{E_c^n}\sigma_0 E e^{-E_c/|E|} \qty(\frac{E_c}{E})^{2n}  
\end{align}
Therefore, for a fixed $E < E_c$, the $n$th derivative in a unit of $J_0/E_c^n$ diverges exponentially with $n$.
Such a rapid change of $J(E)$ is a unique feature arising from the quantum power threshold, and significantly enhances the high harmonic transport in quantum insulators.  

\subsection{Properties of the modified Bessel function of the second kind $K_{\nu}(z)$} \label{ap:subsec:Knu}

A useful expression for the modified Bessel function of the second kind $K_{\nu}(z)$ in terms of an integral is~\cite{NIST:DLMF}
\begin{align}
    \int_0^\infty x^{\nu-1}
    \exp\!\left(-\frac{\beta}{x}-\gamma x\right)dx
    =2\left(\frac{\beta}{\gamma}\right)^{\nu/2}
    K_\nu\!\left(2\sqrt{\beta\gamma}\right). \label{ap:eq:Knu}
\end{align}
This is valid for $\Re\beta>0$ and $\Re\gamma>0$.

The asymptotic behavior of $K_{\nu}(z)$ is the following. For $\Re\nu>0$, $K_{\nu}(z)$ behaves at $z\to 0$ as 
\begin{align}
    K_{\nu}(z\to0) = \frac{2^{\nu-1}\Gamma(\nu)}{z^{\nu}} + \dots \label{eq:K_nu_zero}
\end{align}
For $\nu=2$, we show here the next leading order for $K_2(z)$ at small $z$ since it is used in deriving the asymptotic behavior for $d=1$ singular high-harmonic transport:
\begin{align}
    K_2(z) = \frac{2}{z^{2}} - \frac{1}{2} + \order{z^2\log z}, \label{ap:eq:K2}
\end{align}
On the other hand, when $|z|\to \infty$ with $|\arg{z}| < 3\pi/2$, 
\begin{align}
    K_{\nu}(z \to \infty) = \sqrt{\frac{\pi}{2z}}e^{-z} \label{ap:eq:Knu_inf}
\end{align}

\end{widetext}

\end{document}